\documentclass[prd,preprint,showpacs,floatfix,amsmath,amssymb,floatfix]{revtex4}
\usepackage{graphicx,color,dcolumn,booktabs,bm}
\usepackage{longtable,lscape}
\usepackage{txfonts}
\usepackage{overpic}
\usepackage{amssymb}
\usepackage{indentfirst}
\usepackage{feynmf}   
\usepackage{slashed}  
\usepackage{cases}
\usepackage{color}
\usepackage{multirow}
\usepackage{graphicx,color,dcolumn,booktabs,bm}
\usepackage{epsfig,dsfont,amssymb,amsmath,amsfonts,amsbsy,mathrsfs}
\usepackage[colorlinks, citecolor=blue,anchorcolor=red,menucolor=red, linkcolor=red,filecolor=red,runcolor=red,urlcolor=blue,frenchlinks=red]{hyperref}

\graphicspath{{Figures/}} %

\begin{document}

\title{A possible global group structure for exotic states }

\author{Xue-Qian Li$^1$}\email{lixq@nankai.edu.cn}
\author{Xiang Liu$^{2,3}$}\email{xiangliu@lzu.edu.cn}
\affiliation{$^1$School of Physics, Nankai University, Tianjin 300071, China\\
$^2$School of Physical Science and Technology, Lanzhou University,
Lanzhou 730000, China\\
$^3$Research Center for Hadron and CSR Physics,
Lanzhou University $\&$ Institute of Modern Physics of CAS,
Lanzhou 730000, China}

\begin{abstract}
\noindent

Based on the fact that the long expected pentaquark which possesses  the exotic quantum numbers
of $B=1$ and $S=1$ was not experimentally found, although exotic states of $XYZ$ have been observed recently, we conjecture that the heavy flavors
may play an important role in stabilizing the hadronic structures beyond the traditional $q\bar q$ and $qqq$
composites.
\end{abstract}
 \maketitle

The success of the $SU(3)$ quark model \cite{GellMann:1964nj,Zweig:1981pd,Zweig:1964jf}, by which we understand the baryons and mesons as composites of quarks,  is a victory of 20th century. However, it definitely is not the end of the story; people are tempted to
ask if the most economic quark structures, where mesons are made of a quark and an antiquark whereas baryons consist of three quarks, are complete.
The discovery of the exotic states of hadrons is undoubtedly an important breakthrough in our understanding of the quark structure and non-perturbative QCD,
which is responsible for holding quarks and gluons inside the color-singlet hadrons.

Later, it was conjectured that some exotic states, such as glueballs \cite{Gui:2012gx,Yang:2013xba,Chen:2014iua,Gao:1995ni,Klempt:2007cp,Crede:2008vw}, which are made of only gluons; hybrids \cite{Klempt:2007cp,Meyer:2010ku,Ke:2007ih}, which contain quarks and gluons, exist;  in the string picture of hybrids
where the gluon degree of freedom is excited, multi-quark states may exist as exotic states. Based on  phenomenological studies, it is suggested that hadrons \cite{Golowich:1982kx,Golowich:1978gj,Gaillard:1985km,Zou:2009zz,Zou:2010tc} may possess such multi-quark components as
$$|B\rangle=C_1|qqq\rangle+C_2|qqqq^\prime\bar q^\prime\rangle+C_3|qqqg\rangle+\cdot\cdot\cdot,$$
and
$$|M\rangle=C^\prime_1|q\bar q\rangle+C^\prime_2|q\bar q g\rangle+C^\prime_3|q\bar qq^\prime\bar q^\prime\rangle+\cdot\cdot\cdot,$$
where the ellipses denote higher order terms of the same quantum numbers. Usually the extra terms are called exotic components.

Such exotic states not only exist as components but also can appear as independent hadrons.
An obvious indication of their existence would be discovering exotic hadrons which have quantum numbers
that the simplest quark version cannot accommodate. For example, if the $1^{-+}$ meson is observed, it definitely is not a regular meson. Moreover, it was suggested
to search for baryons whose baryon number is +1 and whose strangeness is also +1 \cite{Gao:1999ar}, this baryon, the  so-called pentaquark, cannot be interpreted as a three-quark structure. Thus the $1^{-+}$ meson must be a hybrid or four-quark state whereas the pentaquark of $B=1$ and $S=1$ at least is in the structure of
$|qqq d \bar{s}\rangle$. Such structures are beyond the simplest singlet, octet, or decuplet representations of $SU(3)$.

The $SU(3)$ quark model surely does not forbid such exotic states. But unfortunately, after a long time search for pentaquarks which was one
of the hottest topics for experimentalists and theorists of high energy physics a decade ago,  all efforts failed. It was announced that the pentaquark
does not exist. Is that really true?  One would ask why exotic states are not experimentally observed as they are predicted by a beautiful theory. Is there some
unknown factor resulting in this phenomenon?

Recently, at the meson sector, many, so called $XYZ$ particles \cite{Liu:2013waa,Chen:2013wva,Brambilla:2010cs,Brambilla:2014jmp} have been observed by the CLEOc, BaBar, Belle, CDF, D0, BESIII, LHCb, and CMS collaborations.
Most of them seem not to be regular mesons, but exotic states by their production and decay characteristics. Those newly discovered mesons have one common
point, i.e., they all contain a hidden heavy flavor $b$ or $c$. Namely they are in the structure of $b\bar b (c\bar c)q\bar q^\prime$.

This observation motivates us to conjecture that the heavy flavor might play an important role to stabilize the multi-quark structure. One is easily to be convinced that with the
heavy flavor the multi-quark mesons are much more stable than those possible multi-quark hadrons containing only light flavors. Indeed, we can find evidence to
strongly support this idea.

The hydrogen molecule which contains two protons and two electrons are rather stable. Based on the Born-Oppenheimer approximation \cite{HO},
Heitler and London calculated the spectrum of the hydrogen molecule which is well consistent with the data. It is a stable structure. Recently,
the positronium molecule was also experimentally observed \cite{Nature,Puchalski:2007ck,Puchalski:2008jj} right after the discovery of the positronium ion, which consists of two electrons and one
positron (or two positrons and one electron). By contrast to the hydrogen molecule, the positronium molecule which consists of two electrons and two
positrons in close analogy to the hydrogen molecule are hard to produce and rather unstable. The situation of positronium ion is similar;  comparing with
a hydrogen ion with two protons and an electron, it is unstable. It is a hint that the two protons, which are 2,000 times heavier than electrons, are responsible for stabilizing the
structure. By analogy, we may conjecture that the two heavy quarks (indeed, one heavy quark and one heavy antiquark)  stabilize the exotic structure.

Generally, for the positronium, annihilation of $e^+$ and $e^-$ might induce effects influencing the inner structure of positronium, namely it varies the effective potential between the electron and positron
and the resultant spectrum might be affected. Moreover, the electron and positron may annihilate into two photons and it would be the main decay mode of the positronium ion and molecule. Instead, the 
electron-proton  annihilation is rare (it is forbidden in the Standard Model) and generally is ignored in atomic physics. The hydrogen atom or the molecule dissolves via absorbing  photons from environment, so its lifetime is rather long; instead, due to
the annihilation, the lifetime of the positronium ion and molecule is short. It is not the concern of this work.
Indeed, the annihilation of an electron-positron pair may affect the stability of the positronium molecule, but there is another reason that the reduced mass for positronium is half
that for hydrogen atom, so the effective radius is doubled, and the same situation may be extended to the case of the postitronium molecule and the hydrogen molecule. The reduced mass is larger when one of the constituents is heavier. We do not fully attribute all the instability to the heavy flavor, but we indicate the tendency.  

We first consider that the heavy constituents in the hadron to really play a role to stabilize the structure based on naive observations. That was indeed an assumption, but it was not
derived from some underlying theories. We use the comparison between positronium and hydrogen molecule as evidence to support this assumption; definitely it is not a strong
evidence.

Very recently, Brodsky {\it et al.} published a paper about the exotic $XYX$ states \cite{Brodsky:2014xia}. They argued "Why are tetraquark states obvious in the charm (and likely bottom) sector but not the light-quark sector?". Again, we believe that our picture provides guidance based on the energy scales of the tetraquark stakes. They analyzed the tetraquark states, and they determine that the heavy constituents indeed are important for the
stabilization of multi-quark states.

There exists a main difference between the two cases that the interaction between proton and electron, or between proton and proton, electron and electron
is the Coulomb potential which is clear and very familiar to researchers, but by contrast, the interactions among quarks  are much more complicated.
The interaction among quarks and gluons is described by QCD, but at lower energies which are the typical scale for binding quarks into hadrons and have the order of $\Lambda_{QCD}$,
the coupling is large and the perturbative theory would no longer work at all. Unfortunately, the non-perturbative QCD causes much trouble for evaluating the hadron
properties including the mass spectra and wavefunctions. Concrete phenomenological models, such as the Cornell potential model, are invoked to deal with the hadron's inner structure.
In the model, besides the Coulomb-type term caused by the one-gluon exchange, a linear term appears to be responsible for the quark confinement, which is completely due to the non-perturbative QCD and could not be derived from any basic theory so far. Therefore, the effective interactions between quarks and gluons inside hadrons are not
clear. However, just as the $SU(3)$ quark model was just established where only the group structure was considered and the relations as regards the mass spectrum of the octet
baryons were exactly obtained and, moreover, the mass of $\Omega$  baryon of the decuplet was first predicted. Therefore, we are tempted to  generalize the group structure for
four-quark states.

Based on our argument given above, the involvement of heavy quarks stabilizes the four-quark states, their existence is necessary. The generalized group
is the global $$G=SU_c(3)\times SU_H(2)\times SU_L(3),$$
where the subscripts $c$, $H$, and $L$ refer to color, heavy, and light, respectively.  The $SU_L(3)$ corresponds to the regular quark model for the light quarks $u,d,s$
and the newly introduced $SU_H(2)$ involves $c$ and $b$ quarks (antiquarks). This idea is inspired by the heavy quark effective theory (HQET) \cite{Isgur:1991xa,Manohar:2000dt}.
Any hadron states must be in a color singlet, so four-quark states can be classified into  the molecular and tetraquark  states according to their color structures. Instead the pentaquark-type baryons would have slightly more complicated structures corresponding to various color combinations.
For $SU_H(2)$ the doublet is
\begin{equation}
H_1=
\left( \begin{array}{c} c \\ b \end{array}\right),
\end{equation}
and its conjugate is $H_2=\left( \begin{array}{c} \bar b \\ \bar c \end{array} \right)$. That is in analogous to the strong isospin doublet
where the doublet is $\left( \begin{array}{c} u \\ d \end{array} \right)$ whose conjugate is $\left( \begin{array}{c} \bar d \\ \bar u \end{array} \right)$.
The $SU_H(2)$ triplet and singlet are, respectively,
$$c\bar b,\; (c\bar c+b\bar b)/\sqrt 2,\; b\bar c,\;\;\;({\rm triplet}),$$
$$(c\bar c-b\bar b)/\sqrt 2, \;\;\; (\rm singlet).$$

For convenience,
we can write $\bar bb,\; \bar cc,\; (\bar bc+\bar c b)/\sqrt 2,\; (\bar bc-\bar c b)/\sqrt 2$ corresponding to
$H_2^T(\sigma_x-i\sigma_y)/2)H_1$, $H_2^T(\sigma_x+i\sigma_y)/2)H_1$, $H_2^T(I/\sqrt 2)H_1$ and $H_2^T(\sigma_z/\sqrt 2)H_1$, respectively,
where $\sigma_i\; (i=x,y,z)$ are the Pauli matrices and $I$ is the unit matrix.

According to the group structure, if the four-quark state resides in an $SU_L(3)$ singlet or octet and an $SU_H(2)$ singlet, it would be the case for the exotic mesons which
were recently observed as $Z_b(10610)$ \cite{Belle:2011aa}, $Z_b(10650)$ \cite{Belle:2011aa}, $Z_c(3900)$ \cite{Ablikim:2013mio,Liu:2013dau}, $Z_c(4020)$ \cite{ Ablikim:2013wzq} $Z_c(4025)$ \cite{Ablikim:2013emm}, etc.
Moreover, we may expect that it is in a triplet of $SU_H(2)$ and in any of the representations $(15, 6, 3_1, 3_2)$ of $3_H\otimes(3\otimes \bar{3})_L$ for  $SU_L(3)$.
Considering the color-singlet requirement, the possible group representations can be $(1, 2, (15,6, 3_1,3_2)$ for $SU_c(3)\times SU_H(2)\times SU_L(3)$ or
possible combinations.

For the baryon case, there are much more possible representations  and,  combining the color requirement, the analysis  becomes very complicated. We will present
a full analysis in our next work.

An alternative possibility is that the hadron is in the other components of the triplet or singlet of $SU_H(2)$. Indeed we have reason to believe that such structures are more stable than the hadrons being in a doublet of $SU_H(2)$. We draw  such a conclusion based on an analogy to the EM case. The hydrogen ion $H_-$ which is also called the hydrogen anion, is resolved by absorbing an energy of 0.75 eV; instead, to resolve a hydrogen molecule into two hydrogen atoms, one needs to provide 4.45 eV.

Therefore, we can understand why the so far observed four-quark states all contain hidden charm or bottom, i.e. the charged $c\bar cu\bar d $ or $b\bar b u\bar d$ and some others \cite{Sun:2012sy}. Based on this conjecture, we can predict the existence of charged $b\bar c q\bar q$.

Another observation is that the exotic states which are made of only light flavors must reside in mixtures with regular hadrons containing $q\bar q$ or $qqq$ \cite{Close:2005vf,He:2006tw,He:2006ij}
to get stabilized. For the supposed pentaquark $uud d\bar s$ unfortunately there does not exist a corresponding regular baryon state, therefore it does not appear or is 
very unstable to evade detection.

Therefore, we would predict that the pentaquarks should be $c\bar c qqq$ and $b\bar b qqq$. However, such baryons would have the same quantum numbers as the regular baryons, unlike their mass spectra, and it is hard to identify them as an exotic state. By contrast, the pentaquark $b\bar c qqq$ \cite{Chen:2014mwa} would have very distinctive quantum numbers as compared to the regular three-quark baryons. Therefore, we suggest to search for such baryons at LHCb whose energy is enough to provide the production phase space.

As a short discussion, we suggest a global group $SU_c(3)\times SU_H(2)\times SU_L(3)$ which can accommodate the observed exotic four-quark states. By assumption, we predict the existence of four-quark states  which contain only one heavy quark, but such states are less stable than that containing hidden heavy flavors. We also suggest the existence of 
$c\bar b(b\bar c)q\bar q$ and the pentaquark $b\bar c qqq$ which can be clearly identified experimentally.

The deuteron in principle can also be thought of as a six-quark system, but as discussed by Weinberg in Ref. \cite{Weinberg:1965zz}, the deuteron
is not an elementary particle, but a composite. Therefore, unlike the exotic states (maybe we also need to include the newly discovered $d^*$ state \cite{Adlarson:2011bh,Adlarson:2014pxj}), we would account the deuteron or even triton not elementary particles, but attribute them to  another level of physics: nuclei.

In general, people assume these observed $XYZ$ states as molecules of $D\bar{D}^*$, $D^*\bar{D}^*$ etc. because the sums of the constituents $D$ and $D^*$ or $D^*$ and $D^*$ are close to the mass of the hadrons, but it is noted that, in most cases, the sum of the constituents is smaller than the mass of the corresponding hadron. The interaction between the constituents of the molecule is via exchanging light mesons (pion, rho, sigma etc.) and the binding energy calculated in the chiral theory is negative; therefore the pure molecular structures for the mesons are not
preferable. Brodsky {\it et al.} \cite{Brodsky:2014xia} are inclined to consider the $XYZ$ mesons as tetraquarks instead of molecules, whereas we used to suggest them to be mixtures of molecules and tetraquarks. That is different from the deuteron.

The newly observed $d^*$ \cite{Adlarson:2011bh,Adlarson:2014pxj} resonance, which seem to contain six quarks with baryon number being 2 \cite{Huang:2014aca,Huang:2014kja}, composes a challenge to the assumed principle. Unless its lifetime is very short,
our postulated principle might be overthrown. We need further and more accurate experimental measurements on $d^*$ and similar multi-quark states to confirm or negate our assumption.

\vspace{1cm}

\section*{Acknowledgement} We sincerely thank Prof. Jean-Marc Richard from the University of Lyon; the idea of this work is partially motivated by his seminar given at Nankai University
a few years ago. This work is supported by the National Natural Science Foundation of China  under Grants  No. 11075079, No. 11175051, No. 11222547, No. 11175073, and No. 11035006, the Ministry of Education of China (SRFDP under Grant No. 2012021111000), the Fok Ying Tung Education Foundation (Grant No. 131006)

\end{document}